\documentstyle[preprint,tighten,aps,epsfig]{revtex}

\begin{document}
\title{Strange tribaryons}
\author{T. Fern\'andez-Caram\'es$^{(1)}$, A. Valcarce$^{(2)}$, 
H. Garcilazo$^{(3)}$, and P. Gonz\'alez$^{(1)}$}
\address{$(1)$ Dpto. de F\' \i sica Te\'orica and IFIC\\
Universidad de Valencia - CSIC, E-46100 Burjassot, Valencia, Spain}
\address{$(2)$ Grupo de F{\'\i}sica Nuclear and IUFFyM\\
Universidad de Salamanca, E-37008 Salamanca, Spain}
\address{$(3)$ Escuela Superior de F\'\i sica y Matem\'aticas, \\
Instituto Polit\'ecnico Nacional, Edificio 9,\\
07738 M\'exico D.F., Mexico}
\maketitle

\begin{abstract}
We use two-body potentials derived from a constituent quark
cluster model to analyze the bound-state problem of the $\Sigma NN$ system.
The observables of the two-body subsystems, $NN$ and $\Sigma N$, are well 
reproduced. We do not find $\Sigma NN$ bound states, but there are
two attractive channels with a resonance close above the three-body
threshold. These channels are the $(I,J)=(1,1/2)$ and $(0,1/2)$, their
quantum numbers, widths and energy ordering
consistent with the recently measured
strange tribaryons from the $^{4}{\rm He}(K_{{\rm stopped}}^{-},N)$
reactions in the KEK PS E471 experiment.
\end{abstract}

\vspace*{2cm} \noindent Pacs: 13.75.Ev,12.39.Jh,21.45.+v

\newpage

\section{Introduction}

Recently an exotic tribaryon resonance $S^{0}(3115)$ has been measured
using the stopped $K^{-}$ absorption experiment, 
$^4 {\rm He} (K_{\rm stopped}^{-},p)$, at KEK-PS \cite{Suz04}. 
The proton energy distribution shows a monoenergetic peak, 
with a significance of $13\sigma $,
interpreted as the formation of a new kind of neutral tribaryon with isospin 
$I=1$ and strangeness $S=-1$. The extracted mass and width of the
state are $3117.0_{-4.4}^{+1.5}$ MeV and $<$ 21 MeV, respectively, and
its main decay mode is found to be $\Sigma NN$. The recent detailed
analysis of the neutron spectrum in 
$^4 {\rm He} (K_{\rm stopped}^{-},n)$ \cite{Suz05} 
has made manifest a second monoenergetic
peak assigned to the formation of another strange tribaryon $S^{+}(3140)$ 
with a significance of $3.7\sigma $. The mass and width of this
state are deduced to be 
$3140.5_{-0.8}^{+3.0}({\rm syst})\pm 2.3({\rm stat})$ MeV
and $<$ 21.6 MeV, respectively, its main decay mode being $\Sigma ^{\pm }NN$.
The isospin of the state is assigned to be zero. The experimental
determination of spin-parity of these strange tribaryons is awaited.

These experimental studies were motivated by the theoretical prediction
of two deeply bound states in the $K^{-} \, ^{3}{\rm He}$ system 
\cite{Aka02}. However neither their predicted binding energies 
nor their isospin level ordering correspond to the measured 
tribaryons. Actually a $I=0$ state was predicted as the deepest 
one with a mass of about 3195 MeV (the $I=1$ state being 87 MeV above).
These difficulties may be resolved by taking care of
relativistic effects and by invoking an enhanced $\overline{K}N$ interaction
and a strong spin-orbit interaction in the dense nuclear medium \cite{Aka05}.
A similar isospin reversing problem is found with the results of
the $SU(3)$ multiskirmion description of multibaryon systems \cite{Sca00}. 
This model predicts the $B=3$ and $S=-1$ lighter resonance to be a 
$(I,J^{\pi })=(0,1/2^{+})$ with an $I=1$ excited state 40 MeV higher, 
both belonging to the $35^*-$plet of flavor. 
Very recently,
a nona-quark interpretation of the strange tribaryons has been suggested,
identifying the $S^{0}(3115)$ as a member of flavor $27-$plet with 
$(I,J)=(1,1/2)$ or $(1,3/2)$ and the $S^{+}(3140)$ as a member of flavor
$10^*-$plet with $(I,J)=(0,3/2)$ or of flavor $35^*$-plet with 
$(I,J)=(0,1/2)$ \cite{Mae05}.

In this work we study the possible existence of $\Sigma NN$ 
positive parity bound states using two-body potentials 
derived from a constituent quark cluster model.
For this purpose we follow the same procedure that we
used in the past to study three-body systems made of $N^{\prime }s$ and 
$\Delta ^{\prime }s$. The three-body
calculations are performed using a truncated $T-$matrix approximation where
the inputs of the three-body equations are the two-body $t-$matrices truncated
such that the orbital angular momentum in the initial and final states is
equal to zero. These two-body $t-$matrices, however, have been constructed
taking into account the coupling to the $\ell=2$ states due to the tensor
force. This approximation in the case of the three-nucleon system, with the
$NN$ interaction taken as the Reid soft-core potential, leads to a triton
binding energy which differs less than 1 MeV from the exact value 
\cite{Har72,Ber86}. In a first approach, 
like in our previous studies of the bound-state 
problem of the $\Delta NN$, $\Delta \Delta N$, and $\Delta \Delta \Delta$ 
systems \cite{Gar97,Mot99,Gar99}, 
we deal with real integral equations 
since we do not consider the imaginary terms arising from the coupling 
of baryon-baryon subsystems to lower mass channels, i.e.,
from the coupling of the $\Sigma N$ subsystem to 
the $\Lambda N$ channel. Then, in a second more complete study we 
consider the full $\Sigma NN - \Lambda NN$ system to check the
effect of the coupling to $\Lambda$ channels at
the three-body level.

We use as basic framework for the baryon-baryon interactions the local
potentials obtained from the constituent quark cluster model since this
provides a consistent and universal treatment for all of them \cite{Val05}.
The paper is organized as follows. In Sec. II we provide a brief description
of the constituent quark model and the formalism to study the two and
three-body systems. In Sec. III we present and discuss our results.
Finally we summarize our main conclusions in Sec. IV.

\section{Formalism}

\subsection{The two-body interactions}

The baryon-baryon interactions involved in the study of the $\Sigma NN$
system are obtained from the constituent quark
cluster model \cite{Val05,Vij05}. In this model baryons
are described as clusters of three interacting massive (constituent) quarks,
the mass coming from the spontaneous breaking of chiral symmetry. The
ingredients of the quark-quark interaction are confinement ($CON$),
one-gluon ($OGE$), one-pion ($\pi $), one-sigma ($\sigma $), one-kaon ($K$)
and one-eta ($\eta $) exchange terms. Explicitly, the quark-quark
interaction reads: 
\begin{equation}
V_{qq}(\vec{r}_{ij})=V_{CON}(\vec{r}_{ij})+V_{OGE}(\vec{r}_{ij})+V_{\pi }(%
\vec{r}_{ij})+V_{\sigma }(\vec{r}_{ij})+V_{K}(\vec{r}_{ij})+V_{\eta }(\vec{r}%
_{ij})\,\,,
\end{equation}%
where the $i$ and $j$ indices are associated with $i$ and $j$ quarks
respectively, ${\vec{r}}_{ij}$ stands for the interquark distance and 
\begin{equation}
V_{CON}({\vec{r}}_{ij})=-a_{c}\,{\vec{\lambda ^{c}}}_{i}\cdot {\vec{\lambda
^{c}}}_{j}\,r_{ij}\,\,,
\end{equation}%
\begin{equation}
V_{OGE}({\vec{r}}_{ij})={\frac{1}{4}}\,\alpha _{s}\,{\vec{\lambda ^c}}_{i}
\cdot {\vec{\lambda ^c}}_{j}\Biggl \lbrace{\frac{1}{r_{ij}}}-
{\frac{1}{4}}\left(
{\frac{1}{{2\,m_{i}^{2}}}}\,+{\frac{1}{{2\,m_{j}^{2}}}}\,+{\frac{2}{%
3m_{i}m_{j}}}{\vec{\sigma}}_{i}\cdot {\vec{\sigma}}_{j}\right) \,\,{\frac{{%
e^{-r_{ij}/r_{0}}}}{{r_{0}^{2}\,\,r_{ij}}}}\Biggr \rbrace\,\,,  \label{reg}
\end{equation}%
\begin{eqnarray}
V_{\pi }({\vec{r}}_{ij}) &=&{\frac{1}{3}}{\frac{g_{ch}^{2}}{{4\pi }}}\,{%
\frac{m_{\pi }^{2}}{{4m_{i}m_{j}}}}\,{\frac{\Lambda _{\chi }^{2}}{\Lambda
_{\chi }^{2}-m_{\pi }^{2}}}\,m_{\pi }\,\left\{ \left[ \,Y(m_{\pi }\,r_{ij})-{%
\frac{\Lambda _{\chi }^{3}}{m_{\pi }^{3}}}\,Y(\Lambda _{\chi }\,r_{ij})%
\right] {\vec{\sigma}}_{i}\cdot {\vec{\sigma}}_{j}\right.  \nonumber \\
&+&\left. \left[ H(m_{\pi }\,r_{ij})-{\frac{\Lambda _{\chi }^{3}}{m_{\pi
}^{3}}}H(\Lambda _{\chi }\,r_{ij})\right] S_{ij}\right\} \sum_{a=1}^{3}{%
(\lambda _{i}^{a}\cdot \lambda _{j}^{a})}\,,  \label{OPE}
\end{eqnarray}%
\begin{equation}
V_{\sigma }({\vec{r}}_{ij})=-{\frac{g_{ch}^{2}}{{4\pi }}}\,{\frac{\Lambda
_{\chi }^{2}}{\Lambda _{\chi }^{2}-m_{\sigma }^{2}}}\,m_{\sigma }\,\left[
Y(m_{\sigma }\,r_{ij})-{\frac{\Lambda _{\chi }}{{m_{\sigma }}}}\,Y(\Lambda
_{\chi }\,r_{ij})\right] \,,  \label{OSE}
\end{equation}%
\begin{eqnarray}
V_{K}(\vec{r}_{ij}) &=&{\frac{1}{3}}{\frac{g_{ch}^{2}}{{4\pi }}}{\frac{%
m_{K}^{2}}{{\ 4m_{i}m_{j}}}}{\frac{\Lambda _{K}^{2}}{{\Lambda
_{K}^{2}-m_{K}^{2}}}}m_{K}\left\{ \left[ Y(m_{K}\,r_{ij})-{\frac{\Lambda
_{K}^{3}}{m_{K}^{3}}}Y(\Lambda _{K}\,r_{ij})\right] (\vec{\sigma}_{i}\cdot 
\vec{\sigma}_{j})\right.  \nonumber \\
&+&\left. \left[ H(m_{K}\,r_{ij})-{\frac{\Lambda _{K}^{3}}{m_{K}^{3}}}%
H(\Lambda _{K}\,r_{ij})\right] S_{ij}\right\} \sum_{a=4}^{7}{(\lambda
_{i}^{a}\cdot \lambda _{j}^{a})}\,,
\end{eqnarray}%
\begin{eqnarray}
V_{\eta }(\vec{r}_{ij}) &=&{\frac{1}{3}}{\frac{g_{ch}^{2}}{{4\pi }}}{\frac{%
m_{\eta }^{2}}{{\ 4m_{i}m_{j}}}}{\frac{\Lambda _{\eta }^{2}}{{\Lambda _{\eta
}^{2}-m_{\eta }^{2}}}}m_{\eta }\left\{ \left[ Y(m_{\eta }\,r_{ij})-{\frac{%
\Lambda _{\eta }^{3}}{m_{\eta }^{3}}}Y(\Lambda _{\eta }\,r_{ij})\right] (%
\vec{\sigma}_{i}\cdot \vec{\sigma}_{j})\right.  \nonumber \\
&+&\left. \left[ H(m_{\eta }\,r_{ij})-{\frac{\Lambda _{\eta }^{3}}{m_{\eta
}^{3}}}H(\Lambda _{\eta }\,r_{ij})\right] S_{ij}\right\} \left[ {\rm cos}%
\theta _{P}(\lambda _{i}^{a=8}\cdot \lambda _{j}^{a=8})-{\rm sin}\theta _{P}%
\right] \,,  \nonumber
\end{eqnarray}%
being 
\begin{equation}
Y(x)\,=\,{\frac{e^{-x}}{x}}\,\,\,\,\,;\,\,\,\,\,H(x)\,=\,\Bigl(1+{\frac{3}{x}%
}+{\frac{3}{{x^{2}}}}\Bigr)Y(x)\,.
\end{equation}%
$a_{c}$ is the confinement strength, the ${\vec{\lambda ^{c}}}$'s (${\vec{%
\lambda ^{a}}}$'s) are the $SU(3)$ color (flavor) matrices, $\alpha _{s}$ is
an effective strong coupling constant, $m_{i}$ is the mass of the quark $i$. 
${\vec{\sigma}}_i$ stands for the Pauli spin operator, $g_{ch}$ is the
chiral coupling constant and $\Lambda _{i}$ are cut-off parameters. 
$m_\pi$, $m_\sigma$, $m_K$ and $m_\eta$ are the masses of the exchanged bosons.
The angle $\theta _{P}$ appears as a consequence of considering the physical 
$\eta $ instead the octet one. Finally, 
$S_{ij}$ is the usual quark-tensor operator
$S_{ij} = 3 (\vec{\sigma}_i \, \cdot \, \hat{r}_{ij}) (\vec{\sigma}_j
\, \cdot \, \hat{r}_{ij}) - \vec{\sigma}_i \, \cdot \vec{\sigma}_j $.
The parameters of the model are those of Ref. \cite{Vij05}.

In order to derive the local $NB_1\to NB_2$ interactions 
($B_i=N,\Delta,\Sigma,\Lambda$) from the
basic $qq$ interaction defined above we use a Born-Oppenheimer
approximation. Explicitly, the potential is calculated as follows,

\begin{equation}
V_{NB_1 (L \, S \, T) \rightarrow NB_2 (L^{\prime}\, S^{\prime}\, T)} (R) =
\xi_{L \,S \, T}^{L^{\prime}\, S^{\prime}\, T} (R) \, - \, \xi_{L \,S \,
T}^{L^{\prime}\, S^{\prime}\, T} (\infty) \, ,  \label{Poten1}
\end{equation}

\noindent where

\begin{equation}
\xi_{L \, S \, T}^{L^{\prime}\, S^{\prime}\, T} (R) \, = \, {\frac{{\left
\langle \Psi_{NB_2 }^{L^{\prime}\, S^{\prime}\, T} ({\vec R}) \mid
\sum_{i<j=1}^{6} V_{qq}({\vec r}_{ij}) \mid \Psi_{NB_1 }^{L \, S \, T} ({\vec R%
}) \right \rangle} }{{\sqrt{\left \langle \Psi_{NB_2 }^{L^{\prime}\,
S^{\prime}\, T} ({\vec R}) \mid \Psi_{NB_2 }^{L^{\prime}\, S^{\prime}\, T} ({%
\vec R}) \right \rangle} \sqrt{\left \langle \Psi_{NB_1 }^{L \, S \, T} ({\vec %
R}) \mid \Psi_{NB_1 }^{L \, S \, T} ({\vec R}) \right \rangle}}}} \, .
\label{Poten2}
\end{equation}
In the last expression the quark coordinates are integrated out keeping $R$
fixed, the resulting interaction being a function of the $N-B_i$ distance. The
wave function $\Psi_{NB_i}^{L \, S \, T}({\vec R})$ for the two-baryon system
is discussed in detail in Ref. \cite{Val05}.

\subsection{The $NN$ and $\Sigma N$ subsystems}

If we consider the system of two baryons $N$ and $B$ ($B=N,\Sigma,\Lambda $)
in a relative $S-$state interacting through a potential $V$ that contains a
tensor force, then there is a coupling to the $NB$ $D-$wave so that the
Lippmann-Schwinger equation of the system is of the form

\begin{eqnarray}
t_{ji}^{\ell s\ell^{\prime \prime }s^{\prime \prime }}(p,p^{\prime \prime };E)
&=&V_{ji}^{\ell s\ell^{\prime \prime }s^{\prime \prime }}(p,p^{\prime \prime
})+\sum_{\ell^{\prime }s^{\prime }}\int_{0}^{\infty }{p^{\prime }}%
^{2}dp^{\prime }\,V_{ji}^{\ell s \ell^{\prime }s^{\prime }}
(p,p^{\prime })  \nonumber \\
&&\times {\frac{1}{E-{p^{\prime }}^{2}/2{\bf \mu }+i\epsilon }}%
t_{ji}^{\ell^{\prime }s^{\prime }\ell^{\prime \prime }s^{\prime \prime
}}(p^{\prime },p^{\prime \prime };E),  \label{eq1}
\end{eqnarray}
where $t$ is the two-body amplitude, $j$, $i$, and $E$ are the
angular momentum, isospin and energy of the system, and $\ell s$, 
$\ell^{\prime }s^{\prime }$, $\ell^{\prime \prime }s^{\prime \prime }$ 
are the initial, intermediate, and final orbital angular momentum 
and spin. $p$ 
and $\mu $ are, respectively, the relative momentum and reduced mass of the
two-body system. More precisely, Eq. (\ref{eq1}) is only valid
for the $\Sigma N$ system with isospin $3/2$ and 
the $NN$ system with isospin $0$. For these cases, the
coupled channels of orbital angular momentum and spin
that contribute are given in the first rows of Tables \ref{t1} 
and \ref{t2}, respectively. 

In the case of the $\Sigma N$ system with isospin $i=1/2$, 
the $\Sigma N$ states are coupled to $\Lambda N$ states. Thus, 
if we denote the $\Sigma N$ system as channel $\Sigma$ and the 
$\Lambda N$ system as channel $\Lambda$, instead
of Eq. (\ref{eq1}) the Lippmann-Schwinger equation for 
$\Sigma N$ scattering with isospin $1/2$ becomes 

\begin{eqnarray}
t_{\alpha\beta;ji}^{\ell_\alpha s_\alpha \ell_\beta s_\beta}(p_\alpha,p_\beta;E) & = & 
V_{\alpha\beta;ji}^{\ell_\alpha s_\alpha \ell_\beta s_\beta}(p_\alpha,p_\beta)+
\sum_{\gamma=\Sigma,\Lambda}\sum_{\ell_\gamma=0,2}
\int_0^\infty p_\gamma^2 dp_\gamma
V_{\alpha\gamma;ji}^{\ell_\alpha s_\alpha \ell_\gamma s_\gamma}(p_\alpha,p_\gamma)
 \nonumber \\ & & \times G_\gamma(E;p_\gamma) 
t_{\gamma\beta;ji}^{\ell_\gamma s_\gamma \ell_\beta s_\beta}
(p_\gamma,p_\beta;E);\,\,\,\,\, \alpha, \beta =\Sigma, \Lambda
\label{pup1}
\end{eqnarray}
where $t_{\Sigma \Sigma;ji}$ is the 
$N\Sigma \rightarrow N\Sigma$ scattering amplitude, 
$t_{\Lambda \Lambda ;ji}$ is the 
$N\Lambda \rightarrow N\Lambda$ scattering amplitude, 
and $t_{\Sigma \Lambda ;ji}$ is
the $N\Sigma \rightarrow N\Lambda$ scattering amplitude. 
The propagators $G_{\Sigma}(E;p_\Sigma)$ and $G_{\Lambda }(E;p_{\Lambda})$ 
in Eq. (\ref{pup1}) are given by

\begin{eqnarray}
G_\Sigma(E;p_\Sigma) &= &{2\mu_{N\Sigma}\over k_\Sigma^2-p_\Sigma^2+i\epsilon},
\label{pup2} \\
G_\Lambda(E;p_\Lambda) &= &{2\mu_{N\Lambda}\over 
k_\Lambda^2-p_\Lambda^2+i\epsilon},
\label{pup3}
\end{eqnarray}
with
\begin{equation}
E=k_\Sigma^2/2\mu_{N\Sigma},
\label{pup3p}
\end{equation}
where the on-shell momenta $k_\Sigma$ and $k_\Lambda$ are related by
\begin{equation}
\sqrt{m_{N}^{2}+k_{\Sigma}^{2}}+\sqrt{m_\Sigma^2+k_\Sigma^2}
=\sqrt{m_{N}^{2}+k_{\Lambda}^{2}}+\sqrt{
m_{\Lambda}^{2}+k_{\Lambda}^{2}}.  
\label{forp10}
\end{equation}
We give in Table \ref{t1} the channels $(\ell_\Sigma,s_\Sigma)$ 
and $(\ell_\Lambda, s_\Lambda )$ corresponding to the $\Sigma N$ 
and $\Lambda N $ systems that are
coupled together for the isospin $1/2$ $\Sigma N$ channels.

In the case of the $NN$ system with isospin $1$ we will take into account 
in an analogous manner the
coupling between the $NN$ and $\Delta N$ systems. If we denote the $NN$ system
as channel $N$ and the $\Delta N$ system as channel $\Delta $, then we
shall write, 

\begin{eqnarray}
t_{\alpha \beta ;ji}^{\ell_\alpha s_\alpha\ell_\beta s_\beta}(p_{\alpha
},p_{\beta };E) &=&V_{\alpha \beta ;ji}^{\ell_\alpha s_\alpha \ell_\beta s_\beta
}(p_{\alpha },p_{\beta })+\sum_{\gamma =N,\Delta }\sum_{\ell_\gamma
s_\gamma}\int_{0}^{\infty }p_{\gamma }^{2}dp_{\gamma }\,V_{\alpha \gamma
;ji}^{\ell_\alpha s_\alpha l_\gamma s_\gamma}(p_{\alpha },p_{\gamma })  
\nonumber \\
&&\times G_{\gamma }(E;p_{\gamma })t_{\gamma \beta ;ji}^{\ell_\gamma
s_\gamma \ell_\beta s_\beta}(p_{\gamma },p_{\beta
};E);\,\,\,\,\,\,\,\,\alpha ,\beta =N, \Delta   \label{eq8}
\end{eqnarray}
where $t_{NN;ji}$ is the 
$NN\rightarrow NN$ scattering amplitude, $t_{\Delta \Delta ;ji}$ is the 
$N\Delta \rightarrow N\Delta $ scattering amplitude, and $t_{N\Delta ;ji}$ is
the $NN\rightarrow N\Delta $ scattering amplitude. The propagators 
$G_{N}(E;p_{N})$ and $G_{\Delta }(E;p_{\Delta })$ in Eq. (\ref{eq8}) are
given by

\begin{eqnarray}
G_{N}(E;p_{N}) &=&{\frac{2\mu _{NN}}{k_{N}^{2}-p_{N}^{2}+i\epsilon }}, 
\label{pup2p} \\
G_{\Delta }(E;p_{\Delta }) &=&{\frac{2\mu _{N\Delta }}{k_{\Delta
}^{2}-p_{\Delta }^{2}+i\epsilon }},  \label{eq9}
\end{eqnarray}
with
\begin{equation}
E=k_N^2/2\mu_{NN},
\label{puprp}
\end{equation}
where the on-shell momenta $k_{N}$ and $k_{\Delta }$ are related by
\begin{equation}
2\sqrt{m_{N}^{2}+k_{N}^{2}}=\sqrt{m_{N}^{2}+k_{\Delta }^{2}}+\sqrt{
m_{\Delta }^{2}+k_{\Delta }^{2}}.  \label{for10}
\end{equation}
We give in Table \ref{t2} the channels $(\ell_N,s_N)$ and $(\ell_\Delta,
s_\Delta)$ corresponding to the $NN$ and $\Delta N$ systems that are
coupled together for the isospin $1$ $^{1}S_{0}$ $NN$ channel.

As mentioned before, for the solution of the three-body system we will use
only the component of the $t-$matrix obtained from the solution of 
Eq. (\ref{eq1}) with $\ell=\ell^{\prime \prime}=0$, and of
Eqs. (\ref{pup1}) and (\ref{eq8}) 
with $\ell_\alpha=\ell_\beta=0$. For that purpose we
define the $S-$wave truncated amplitude which in the case of the $\Sigma N$
system with isospin $3/2$ and the $NN$ system with isospin $0$ 
is defined from the solution of Eq. (\ref{eq1}) by 
\begin{equation}
t_{k;si}(p,p^{\prime \prime };E)\equiv t_{ji}^{0s0s^{\prime\prime}}(p,p^{\prime \prime };E)
\,\,\, , \,\,\,\,\,\, k=NN,\Sigma \Sigma \,\, ;
\label{eq7}
\end{equation}
and for the $\Sigma N$-$\Lambda N$ system with isospin 1/2 and the
$NN$ system with isospin 1 is defined, respectively, from the solution of Eqs. 
(\ref{pup1}) and (\ref{eq8}) by 
\begin{equation}
t_{k;si}(p,p^{\prime \prime };E)\equiv t_{\alpha\beta;ji}^{0s_\alpha0s_\beta}(p,p^{\prime \prime
};E)\,\,\, , \,\,\,\,\,\, k=\alpha\beta=NN,\Sigma \Sigma,
\Sigma\Lambda,\Lambda\Sigma,\Lambda\Lambda.
\label{eq7p}
\end{equation}

\subsection{The $\Sigma NN$ system}
\label{II.C}

The numerical solution of the bound-state problem in the case of the $\Sigma
NN$ system will be obtained using the same formalism  used in Ref. 
\cite{Mot99} for the case of the $\Delta \Delta N$ system since in both
cases one is dealing with a system with two identical particles and a third
one which is different. The effects of the $\Lambda N$ and $\Delta N$ 
channels are included in the calculation of the $N\Sigma$ and $NN$
$t-$matrices, respectively, as indicated in Eqs. (\ref{pup1}) and
(\ref{eq8}). Since we are going to apply this formalism to the $\Sigma NN$
bound-state problem the two-body propagators $G_\Sigma$, $G_N$, and
$G_\Delta$ given by Eqs. (\ref{pup2}), (\ref{pup2p}), and (\ref{eq9})
never blow up. Only the propagator $G_\Lambda$ given by Eq. (\ref{pup3})
blows up since the $\Lambda N$ channel is open, so that

\begin{equation}
G_\Lambda(E;p_\Lambda)={2\mu_{N\Lambda}\over k_\Lambda^2-p_\Lambda^2}
-\pi i 2\mu_{N\Lambda}\delta(k_\Lambda^2-p_\Lambda^2).
\label{far1}
\end{equation}
However, from Ref. \cite{Gaj97} we expect the imaginary part of the
propagator $G_\Lambda$ contributing mainly to the width of the $\Sigma NN$
states while having very little effect on their masses. Therefore, 
in order to calculate the masses of the states we will neglect
the imaginary part of this propagator so that our calculations will
be done taking only the real part, i.e.,  

\begin{equation}
G_\Lambda(E;p_\Lambda)\equiv{2\mu_{N\Lambda}\over k_\Lambda^2-p_\Lambda^2}.
\label{far2}
\end{equation}
However, when calculating their widths the full propagator, Eq. (\ref{far1}),
will be used.

If we restrict ourselves to the
configurations where all three particles are in $S-$wave states, the Faddeev
equations for the bound-state problem in the case of three baryons
with total spin $S$ and total isospin $I$ are 
\begin{eqnarray}
T_{i;SI}^{s_{i}i_{i}}(p_{i}q_{i}) &=&\sum_{j\neq
i}\sum_{s_{j}i_{j}}h_{ij;SI}^{s_{i}i_{i}s_{j}i_{j}}{\frac{1}{2}}%
\int_{0}^{\infty }q_{j}^{2}dq_{j}\int_{-1}^{1}d{\rm cos}\theta
\,t_{i;s_{i}i_{i}}(p_{i},p_{i}^{\prime };E-q_{i}^{2}/2\nu _{i})  \nonumber \\
&&\times {\frac{1}{E-p_{j}^{2}/2\mu _{j}-q_{j}^{2}/2\nu _{j}}}%
\,T_{j;SI}^{s_{j}i_{j}}(p_{j}q_{j}),  \label{for1}
\end{eqnarray}%
where $t_{1;s_{1}i_{1}}$ stands for the two-body $NN$
amplitude, and $t_{2;s_{2}i_{2}}$ and $t_{3;s_{3}i_{3}}$ for the 
$\Sigma N$ amplitudes.
$p_{i}$ is the momentum of the pair $jk$ (with $ijk$ an
even permutation of 123) and $q_{i}$ the momentum of
particle $i$ with respect to the pair $jk$. $\mu _{i}$
and $\nu _{i}$ are the corresponding reduced masses

\begin{eqnarray}
\mu _{i} &=&{\frac{m_{j}m_{k}}{m_{j}+m_{k}}},  \label{for2} \\
\nu _{i} &=&{\frac{m_{i}(m_{j}+m_{k})}{m_{i}+m_{j}+m_{k}}} \, .  
\label{for3}
\end{eqnarray}

\noindent
The momenta ${p'}_i$ and $p_j$ in Eq. (\ref{for1}) are given by

\begin{eqnarray}
{p^\prime_i}^2 & = & q_j^2 + {\mu_i^2 \over m_k^2}q_i^2
+ 2{\mu_i \over m_k}q_i q_j {\rm cos}\theta, 
\label{for4} \\
p_j^2 & = & q_i^2 + {\mu_j^2 \over m_k^2}q_j^2
+ 2{\mu_j \over m_k}q_i q_j {\rm cos}\theta. 
\label{for5}
\end{eqnarray}
$h_{ij;SI}^{s_ii_is_ji_j}$ are the spin-isospin coefficients

\begin{eqnarray}
h_{ij;SI}^{s_ii_is_ji_j} & = & (-)^{s_j+\sigma_j-S}\sqrt{(2s_i+1)(2s_j+1)}\,
W(\sigma_j \sigma_k S \sigma_i;s_i s_j) \nonumber \\
& & \times (-)^{i_j+\tau_j-I}\sqrt{(2i_i +1)(2i_j +1)}\,
W(\tau_j \tau_k I \tau_i;i_i i_j), 
\label{for6}
\end{eqnarray}
where $W$ is the Racah coefficient 
and $\sigma_i$, $s_i$, and $S$ ($\tau_i$,
$i_i$, and $I$) are the spins (isospins) of particle $i$, of
the pair $jk$, and of the three-body system.

Since the variable $p_i$ in Eq. (\ref{for1}) [also in Eqs. (\ref{eq1}),
(\ref{pup1}), and (\ref{eq8})] runs from $0$
to $\infty$ it is convenient to make the transformation

\begin{equation}
x_{i}={\frac{p_{i}-b}{p_{i}+b}},  \label{for9}
\end{equation}%
where the new variable $x_{i}$ runs from $-1$ to $1$ and $b$ is a scale
parameter. With this transformation Eq. (\ref{for1}) takes the form

\begin{eqnarray}
T_{i;SI}^{s_ii_i}(x_iq_i) & = & \sum_{j\ne i} \sum_{s_ji_j}
h_{ij;SI}^{s_ii_is_ji_j}{1 \over 2}
\int_0^\infty q_j^2 dq_j \int_{-1}^1 d{\rm cos}\theta\, 
t_{i;s_ii_i}(x_i,x^\prime_i;
E - q_i^2/2\nu_i) \nonumber \\
& & \times {1 \over E - p_j^2/2\mu_j -q_j^2/2\nu_j}\,
T_{j;SI}^{s_ji_j}(x_jq_j).
\label{fo10}
\end{eqnarray}
Since in the amplitude $t_{i;s_ii_i}(x_i,x^\prime_i;e)$ 
the variables $x_i$ and ${x'}_i$
run from $-1$ to $1$, one can expand this amplitude in terms of Legendre
polynomials as

\begin{equation}
t_{i;s_ii_i}(x_i,x^\prime_i;e)=\sum_{nr}P_n(x_i)
\tau_{i;s_ii_i}^{nr}(e)P_r(x^\prime_i),
\label{for11}
\end{equation}
where the expansion coefficients are given by

\begin{equation}
\tau_{i;s_ii_i}^{nr}(e)={2n+1 \over 2}\,{2r+1 \over 2}
\int_{-1}^1 dx_i \int_{-1}^1
dx^\prime_i\, P_n(x_i)t_{i;s_ii_i}(x_i,x^\prime_i;e)
P_r(x^\prime_i).
\label{for12}
\end{equation}
Applying expansion (\ref{for11}) in Eq. (\ref{fo10}) one gets 

\begin{equation}
T_{i;SI}^{s_ii_i}(x_iq_i) = \sum_n T_{i;SI}^{ns_ii_i}(q_i)P_n(x_i),
\label{for13}
\end{equation}
where $T_{i;SI}^{ns_ii_i}(q_i)$ satisfies the one-dimensional integral
equation

\begin{equation}
T_{i;SI}^{ns_{i}i_{i}}(q_{i})=\sum_{j\neq
i}\sum_{ms_{j}i_{j}}\int_{0}^{\infty
}dq_{j}%
\,A_{ij;SI}^{ns_{i}i_{i}ms_{j}i_{j}}(q_{i},q_{j};E)T_{j;SI}^{ms_{j}i_{j}}(q_{j}),
\label{for14}
\end{equation}%
with 

\begin{eqnarray}
A_{ij;SI}^{ns_{i}i_{i}ms_{j}i_{j}}(q_{i},q_{j};E)
&=&h_{ij;SI}^{s_{i}i_{i}s_{j}i_{j}}\sum_{r}\tau
_{i;s_{i}i_{i}}^{nr}(E-q_{i}^{2}/2\nu _{i}){\frac{q_{j}^{2}}{2}}  \nonumber \\
&&\times \int_{-1}^{1}d{\rm cos}\theta \,{\frac{P_{r}(x^\prime_{i})P_{m}(x_{j})}{%
E-p_{j}^{2}/2\mu _{j}-q_{j}^{2}/2\nu _{j}}}.  \label{for15}
\end{eqnarray}

The three amplitudes $T_{1;SI}^{ls_{1}i_{1}}(q_{1})$, 
$T_{2;SI}^{ms_{2}i_{2}}(q_{2})$, and $T_{3;SI}^{ns_{3}i_{3}}(q_{3})$ in Eq.
(\ref{for14}) are coupled together. The number of coupled equations can be
reduced, however, since two of the particles are identical. The reduction
procedure for the case where one has two identical fermions has been
described before \cite{Afn74,Gam90} and will not be repeated here. 
With the assumption that particle 1 is the $\Sigma$ and
particles 2 and 3 are the nucleons, only the amplitudes 
$T_{1;SI}^{ns_{1}i_{1}}(q_{1})$ and $%
T_{2;SI}^{ms_{2}i_{2}}(q_{2}) $ are independent from each other and they
satisfy the coupled integral equations

\begin{eqnarray}
T_{1;SI}^{rs_{1}i_{1}}(q_{1}) &=&2\sum_{ms_{2}i_{2}}\int_{0}^{\infty
}dq_{3}%
\,A_{13;SI}^{rs_{1}i_{1}ms_{2}i_{2}}(q_{1},q_{3};E)T_{2;SI}^{ms_{2}i_{2}}(q_{3}),
\label{for16} \\
T_{2;SI}^{ns_{2}i_{2}}(q_{2}) &=&\sum_{ms_{3}i_{3}}G\int_{0}^{\infty
}dq_{3}%
\,A_{23;SI}^{ns_{2}i_{2}ms_{3}i_{3}}(q_{2},q_{3};E)T_{2;SI}^{ms_{3}i_{3}}(q_{3})
\nonumber \\
&&+\sum_{rs_{1}i_{1}}\int_{0}^{\infty
}dq_{1}%
\,A_{31;SI}^{ns_{2}i_{2}rs_{1}i_{1}}(q_{2},q_{1};E)T_{1;SI}^{rs_{1}i_{1}}(q_{1}) \, ,
\label{for17}
\end{eqnarray}%
with the identical-particle factor

\begin{equation}
G=(-1)^{1+\sigma _{1}+\sigma _{3}-s_{2}+\tau _{1}+\tau _{3}-i_{2}} \, ,
\label{for18}
\end{equation}%
with $\sigma _{1}$ ($\tau _{1})$and $\sigma _{3}$ ($\tau _{3})$ standing for
the spin (isospin) of the $\Sigma $ and the $N$ respectively.

Substitution of Eq. (\ref{for16}) into Eq. (\ref{for17}) yields an equation
with only the amplitude $T_{2}$

\begin{equation}
T_{2;SI}^{ns_2i_2}(q_2) = \sum_{ms_3i_3} \int_0^\infty
dq_3\,K_{SI}^{ns_2i_2ms_3i_3}(q_2,q_3;E) T_{2;SI}^{ms_3i_3}(q_3),
\label{for19}
\end{equation}
where

\begin{eqnarray}
K_{SI}^{ns_2i_2ms_3i_3}(q_2,q_3;E) & = & G
A_{23;SI}^{ns_2i_2ms_3i_3}(q_2,q_3;E)+2\sum_{rs_1i_1} \int_0^\infty dq_1\, 
\nonumber \\
& & \times A_{31;SI}^{ns_2i_2rs_1i_1}(q_2,q_1;E)
A_{13;SI}^{rs_1i_1ms_3i_3}(q_1,q_3;E).  \label{for20}
\end{eqnarray}

In order to find the solutions of Eq. (\ref{for20}) we replace the integral
by a sum applying a numerical integration quadrature \cite{Abra72}. In this
way Eq. (\ref{for20}) becomes a set of homogeneous linear equations. This
set of linear equations has solutions only if the determinant of the matrix
of the coefficients (the Fredholm determinant) vanishes for certain
energies. Thus, the procedure to find the bound states of the system
consists simply in searching for the zeroes of the Fredholm determinant as a
function of energy. We give in Table \ref{t3} the six $\Sigma NN$ states
characterized by total isospin and spin $(I,J)$ that are possible as well as
the two-body $\Sigma N$ and $NN(\Sigma)$ ($NN$ channels with $\Sigma$
spectator) channels that contribute to each state.

Our method of solution of the three-body problem is based in the separable
expansion (\ref{for11}) of the two-body $t-$matrices. We tested in 
Ref. \cite{Mot99} (see Table IV of this reference) the convergence of this
expansion by considering the three-nucleon bound-state problem with the Reid
soft-core potential in the truncated $T-$matrix approximation (two-channel
calculation) \cite{Ber86}. Convergence is reached with 
$N=10$ ($N$ is the number of Legendre polynomials in the separable
expansion) although a very reasonable result is obtained
already with $N=5$. In the calculations of this paper we use $N=10$.

\subsection{The $\Sigma NN - \Lambda NN$ system}
\label{II.D}

The numerical procedure to solve the bound state problem of the 
$\Sigma NN - \Lambda NN$ system is the same as the one described 
in the previous section but considering the full 
propagator $G_\Lambda(E;p_\Lambda)$ in Eq. (\ref{far1}). Besides, when
one includes in addition to the $\Sigma NN$ states also the 
$\Lambda NN$ states, Eq. (\ref{for19}) becomes a two-component
equation, i.e.,

\begin{equation}
T_{2;SI}^{ns_2i_2}(q_2) = \left(\matrix{
T_{2;SI;\Sigma}^{ns_2i_2}(q_2)  \cr
T_{2;SI;\Lambda}^{ns_2i_2}(q_2)  \cr} \right),
\label{gl1}
\end{equation}
and the kernel of Eq. (\ref{for19}) is now a $2\times 2$ matrix defined
by Eq. (\ref{for20}) with

\begin{equation}
A_{23;SI}^{ns_2i_2ms_3i_3}(q_2,q_3;E)=\left(\matrix{
A_{23;SI;\Sigma\Sigma}^{ns_2i_2ms_3i_3}(q_2,q_3;E)&
A_{23;SI;\Sigma\Lambda}^{ns_2i_2ms_3i_3}(q_2,q_3;E)\cr
A_{23;SI;\Lambda\Sigma}^{ns_2i_2ms_3i_3}(q_2,q_3;E)&
A_{23;SI;\Lambda\Lambda}^{ns_2i_2ms_3i_3}(q_2,q_3;E)\cr }\right),
\label{gl2}
\end{equation}

\begin{equation}
A_{31;SI}^{ns_2i_2rs_1i_1}(q_2,q_1;E)=\left(\matrix{
A_{31;SI;\Sigma N(\Sigma)}^{ns_2i_2rs_1i_1}(q_2,q_1;E)&
A_{31;SI;\Sigma N(\Lambda)}^{ns_2i_2rs_1i_1}(q_2,q_1;E)\cr
A_{31;SI;\Lambda N(\Sigma)}^{ns_2i_2rs_1i_1}(q_2,q_1;E)&
A_{31;SI;\Lambda N(\Lambda)}^{ns_2i_2rs_1i_1}(q_2,q_1;E)\cr}\right),
\label{gl3}
\end{equation}

\begin{equation}
A_{13;SI}^{rs_1i_1ms_3i_3}(q_1,q_3;E)  =\left(\matrix{
A_{13;SI;N\Sigma}^{rs_1i_1ms_3i_3}(q_1,q_3;E)  &
   0 \cr
   0&
A_{13;SI;N\Lambda}^{rs_1i_1ms_3i_3}(q_1,q_3;E)  \cr}\right) \, ,
\label{gl4}
\end{equation}
where

\begin{eqnarray}
A_{23;SI;\alpha\beta}^{ns_{2}i_{2}ms_{3}i_{3}}(q_{2},q_{3};E)
&=&h_{23;SI}^{s_{2}i_{2}s_{3}i_{3}}\sum_{r}\tau
_{2;s_{2}i_{2};\alpha\beta}^{nr}(E-q_{2}^{2}/2\nu _{2}){\frac{q_{3}^{2}}{2}}  \nonumber \\
&&\times \int_{-1}^{1}d{\rm cos}\theta \,{\frac{P_{r}(x^\prime_{2})P_{m}(x_{3})}{%
E+\Delta E\delta_{\beta\Lambda}-p_{3}^{2}/2\mu _{3}-q_{3}^{2}/2\nu _{3}
+ i\epsilon}}; \,\,\,\,\,\,\,\, \alpha,\beta=\Sigma,\Lambda,
\label{gl5}
\end{eqnarray}

\begin{eqnarray}
A_{31;SI;\alpha N(\beta)}^{ns_{2}i_{2}ms_{1}i_{1}}(q_{2},q_{1};E)
&=&h_{31;SI}^{s_{2}i_{2}s_{1}i_{1}}\sum_{r}\tau
_{3;s_{2}i_{2;\alpha\beta}}^{nr}(E-q_{2}^{2}/2\nu _{2})
{\frac{q_{1}^{2}}{2}}  \nonumber \\
&&\times \int_{-1}^{1}d{\rm cos}\theta \,{\frac{P_{r}(x^\prime_{3})P_{m}(x_{1})}{%
E+\Delta E\delta_{\beta\Lambda}-p_{1}^{2}/2\mu _{1}-q_{1}^{2}/2\nu _{1}
+ i\epsilon}}; \,\,\,\,\,\,\,\, \alpha,\beta=\Sigma,\Lambda,
\label{gl6}
\end{eqnarray}

\begin{eqnarray}
A_{13;SI;N\beta}^{ns_{1}i_{1}ms_{3}i_{3}}(q_{1},q_{3};E)
&=&h_{13;SI}^{s_{1}i_{1}s_{3}i_{3}}\sum_{r}\tau
_{1;s_{1}i_{1;NN}}^{nr}(E+\Delta E\delta_{\beta\Lambda}-q_{1}^{2}/2\nu _{1})
{\frac{q_{3}^{2}}{2}}  \nonumber \\
&&\times \int_{-1}^{1}d{\rm cos}\theta \,{\frac{P_{r}(x^\prime_{1})P_{m}(x_{3})}{%
E+\Delta E\delta_{\beta\Lambda}-p_{3}^{2}/2\mu _{3}-q_{3}^{2}/2\nu _{3}.  
+ i\epsilon}}; \,\,\,\,\,\,\,\, \beta=\Sigma,\Lambda,
\label{gl7}
\end{eqnarray}
with the isospin and mass of particle 1 (the hyperon) being 
determined by the subindex $\beta$. The subindex $\alpha N(\beta)$ in 
Eq. (\ref{gl6}) indicates a transition $\alpha N \to \beta N$ with a
nucleon as spectator followed by a $NN \to NN$ transition with $\beta$ as
spectator and, 

\begin{equation}
\tau_{i;s_ii_i;\alpha \beta}^{nr}(e)={2n+1 \over 2}\,{2r+1 \over 2}
\int_{-1}^1 dx_i \int_{-1}^1
dx^\prime_i\, P_n(x_i)t_{i;s_ii_i;\alpha \beta}(x_i,x^\prime_i;e)
P_r(x^\prime_i) \, .
\label{for112}
\end{equation}
We give in Table \ref{t3} the six $\Sigma NN - \Lambda NN$ states
characterized by total isospin and spin $(I,J)$ that are possible as well as
the two-body $\Sigma N$, $\Lambda N$, $NN(\Sigma)$ 
($NN$ channels with $\Sigma$ spectator) and 
$NN(\Lambda)$ ($NN$ channels with $\Lambda$ spectator) 
channels that contribute to each state.

The energy shift $\Delta E$, which is usually taken as $M_\Sigma-M_\Lambda$,
will be chosen instead such that at the $\Sigma d$ threshold the momentum
of the $\Lambda d$ system has the correct value in consistency with the
two-body prescription of Eqs. (\ref{forp10}) and (\ref{for10}). Thus, writing

\begin{equation}
E={k_\Sigma^2\over 2\mu_{\Sigma d}},
\label{gl8}
\end{equation}

\begin{equation}
E+\Delta E={k_\Lambda^2\over 2\mu_{\Lambda d}},
\label{gl9}
\end{equation}
where $k_\Sigma$ and $k_\Lambda$ are related by

\begin{equation}
\sqrt{m_{d}^{2}+k_{\Sigma}^{2}}+\sqrt{m_\Sigma^2+k_\Sigma^2}
=\sqrt{m_{d}^{2}+k_{\Lambda}^{2}}+\sqrt{
m_{\Lambda}^{2}+k_{\Lambda}^{2}},
\label{gl10}
\end{equation}
if one takes $E=0$, Eqs. (\ref{gl8})-(\ref{gl10}) lead to

\begin{equation}
\Delta E={[(m_\Sigma+m_d)^2-(m_\Lambda+m_d)^2]
[(m_\Sigma+m_d)^2-(m_\Lambda-m_d)^2]\over
8\mu_{\Lambda d}(m_\Sigma+m_d)^2}.
\label{gl11}
\end{equation}
Since the $\Lambda NN$ channels are in the continuum one has to deal 
with the three-body singularities arising from these channels. Thus,
we used in Eqs. (\ref{for19}) and (\ref{for20}) the 
rotated-contour prescription

\begin{equation}
q_i\to q_i e^{-i\phi};\,\,\,\,\,\,\,\,\, i=1,2,3,
\label{gl12}
\end{equation}
since we found out numerically that the Fredholm determinant does not
depend on the contour-rotation angle $\phi$.

\section{Results}

We will start by presenting the predictions of our model for the $NN$ and 
$\Sigma N$ subsystems and afterwards we discuss the three-body system.

\subsection{The two-body subsystems}

As has been discussed in detail in Ref. \cite{Gar97} a precise
description of the $NN$ low-energy observables is obtained. For the case of
the $^{3}S_{1}-\, ^{3} \! D_{1}$ interaction the model gives the correct binding
energy for the deuteron and a pretty nice description of the phase shifts
(see Figs. 2 and 3 of Ref. \cite{Gar97}). For isospin $1$ channels 
the coupling to the $\Delta N$ system leads to a satisfactory
description of the $NN$ $^{1}S_{0}$ phase shift. The slightly different
tuning of the cut-off for the $^{1}S_{0}$ ($\Lambda _{\chi }=4.38$ fm$^{-1}$)
and $^{3}S_{1}$ ($\Lambda _{\chi }=4.28$ fm$^{-1}$) partial waves
resembles the different value of the $\sigma -$meson parameters used by the Bonn
potential for the same channels, in order to achieve a precise
description of the low-energy data for both partial waves \cite{Mac87}.

We now turn to the available low-energy data on the $\Sigma N$ 
scattering. There is only a small amount of relevant data 
corresponding to the total cross sections (and
some differential cross sections) for $\Sigma
^{+}p\rightarrow \Sigma ^{+}p$, $\Sigma ^{-}p\rightarrow \Sigma ^{-}p$, 
$\Sigma ^{-}p\rightarrow \Sigma ^{0}n$, and $\Sigma^-p\rightarrow\Lambda n$
reactions. It has been known for a long time 
\cite{Swa71} that the available data do not allow for a unique effective
range analysis. This is due (apart from the large error bars) to the absence
of truly low-energy cross sections. The lowest hyperon laboratory
momentum is larger than 100 MeV/c, which means that the inverse
of the scattering length, $1/a$, and the range term, $r k^2/2$, can be of the
same order, leading to results for the scattering length and effective
range that are not unique. This has been clearly illustrated 
in Ref. \cite{Rij99} using six models for the hyperon-nucleon interaction 
with different properties on a detailed level, but providing all of them
with an equally good description of the scattering data.

In the case of processes of the type $\Sigma N\rightarrow\Sigma N$ the
amplitudes obtained from Eqs. (\ref{eq1}) and (\ref{pup1}) are related to 
the effective-range parameters $a$ and $r$ as

\begin{equation}
t_{\Sigma\Sigma;si}^{00}=-{1\over \pi\mu_{N\Sigma}}{1\over 1/a_{si}
+r_{si}k_\Sigma^2/2-ik_\Sigma},
\label{pup4}
\end{equation}
so that the cross section for a given isospin state is

\begin{eqnarray}
\sigma ^{i} &=& \pi^3\mu_{N\Sigma}^2\left(3\mid t_{\Sigma\Sigma;1i}^{00}\mid^2
+\mid t_{\Sigma\Sigma;0i}^{00}\mid^2\right) \nonumber \\
&=& {\frac{{3\pi }%
}{{k_\Sigma^{2}+\left( 1/a_{1i}+r_{1i}k_\Sigma^{2}/2\right) ^{2}}}}+{\frac{{\pi }}{{%
k_\Sigma^{2}+\left( 1/a_{0i}+r_{0i}k_\Sigma^{2}/2\right) ^{2}}}}\,,
\label{pup5}
\end{eqnarray}

We have tuned the interaction to reproduce the different 
total scattering cross sections by using the set of 
parameters of Ref. \cite{Vij05} and 
adjusting the harmonic oscillator parameter of the baryon
wave function. As expected from the calculation
of the root mean square radius of strange baryons \cite{Fur03}
a slightly larger value of $b_s$ is needed ($b_s=0.7$ fm).
From the isospin cross sections (\ref{pup5}) the physical
channels are determined through, 
\begin{eqnarray}
\sigma _{\Sigma ^{+}p} &=&\sigma ^{i=3/2}\,,  \nonumber \\
\sigma _{\Sigma ^{-}p} &=&{\frac{1}{9}}\sigma ^{i=3/2}+{\frac{4}{9}}\sigma
^{i=1/2}\,, \\
\sigma _{\Sigma ^{-}p\rightarrow \Sigma ^{0}n} &=&{\frac{2}{9}}\sigma
^{i=3/2}+{\frac{2}{9}}\sigma ^{i=1/2}\,.  \nonumber
\end{eqnarray}

In the case of the process $\Sigma N\to\Lambda N$ it is necessary to
include also the transition with $\ell=2$ in the $\Lambda N$ channel
since in that channel one is far above threshold. Thus, in that case 
the cross section for isospin $i=1/2$ is

\begin{equation}
\sigma ^{1/2} = \pi^3\mu_{N\Sigma}\mu_{N\Lambda}{k_\Lambda\over k_\Sigma}
\left( \mid t_{\Sigma\Lambda;01/2}^{00}\mid^2
+3 \mid t_{\Sigma\Lambda;11/2}^{00}\mid^2
+3 \mid t_{\Sigma\Lambda;11/2}^{02}\mid^2\right),
\label{pup6}
\end{equation}
and the cross section for the physical channel is 

\begin{equation}
\sigma_{\Sigma^-p\to\Lambda n}={2\over 3}\sigma^{1/2}.
\label{pup7}
\end{equation}

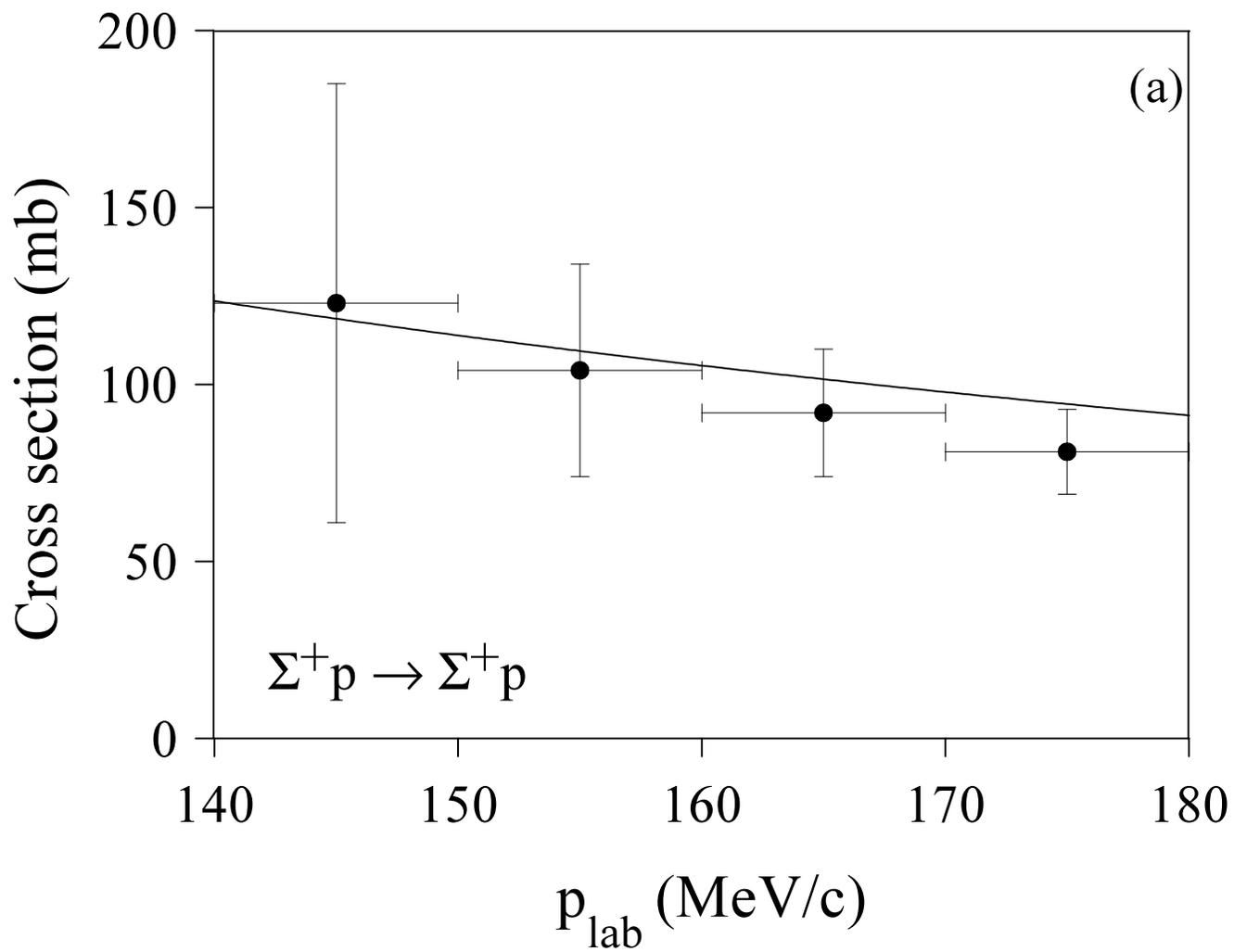
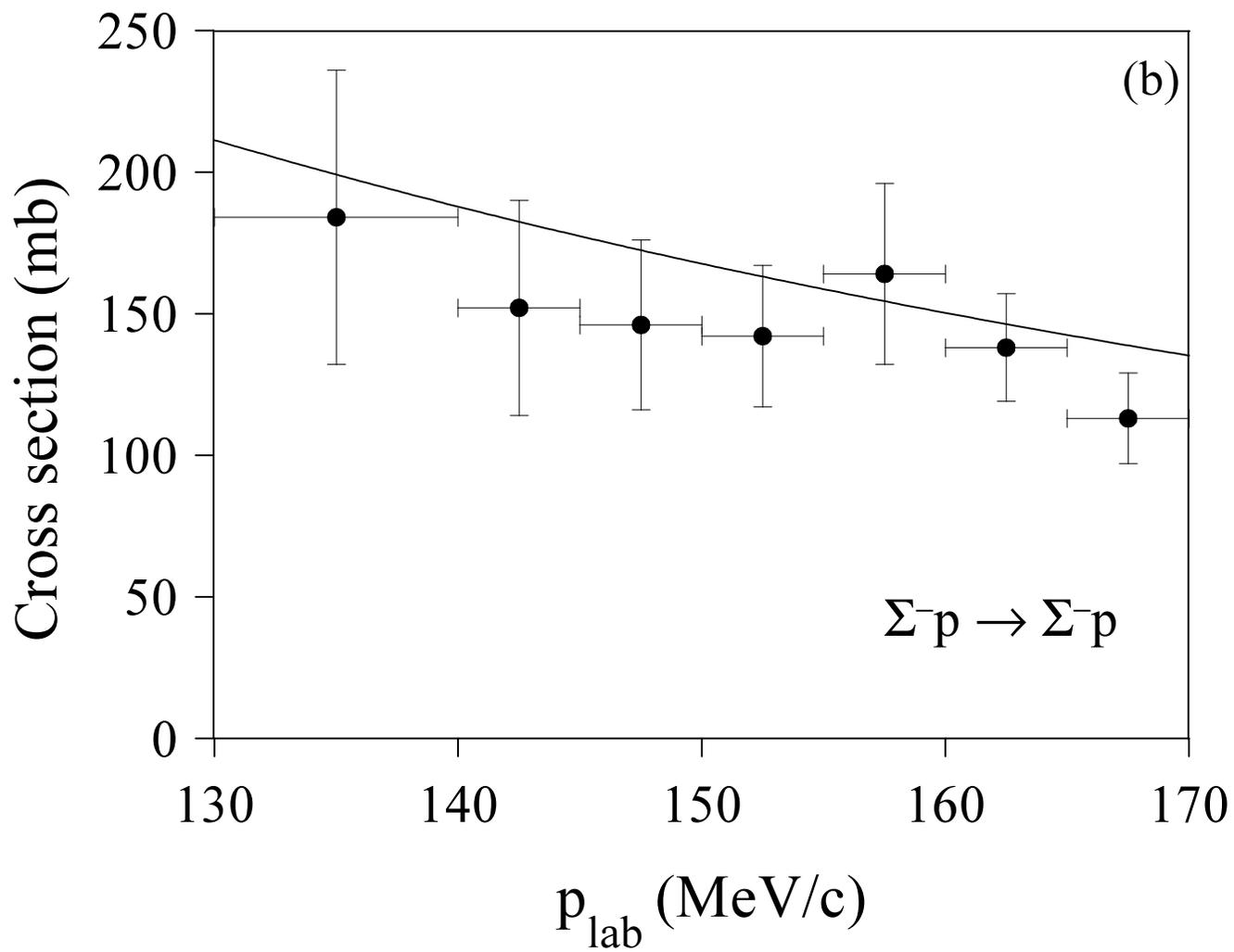
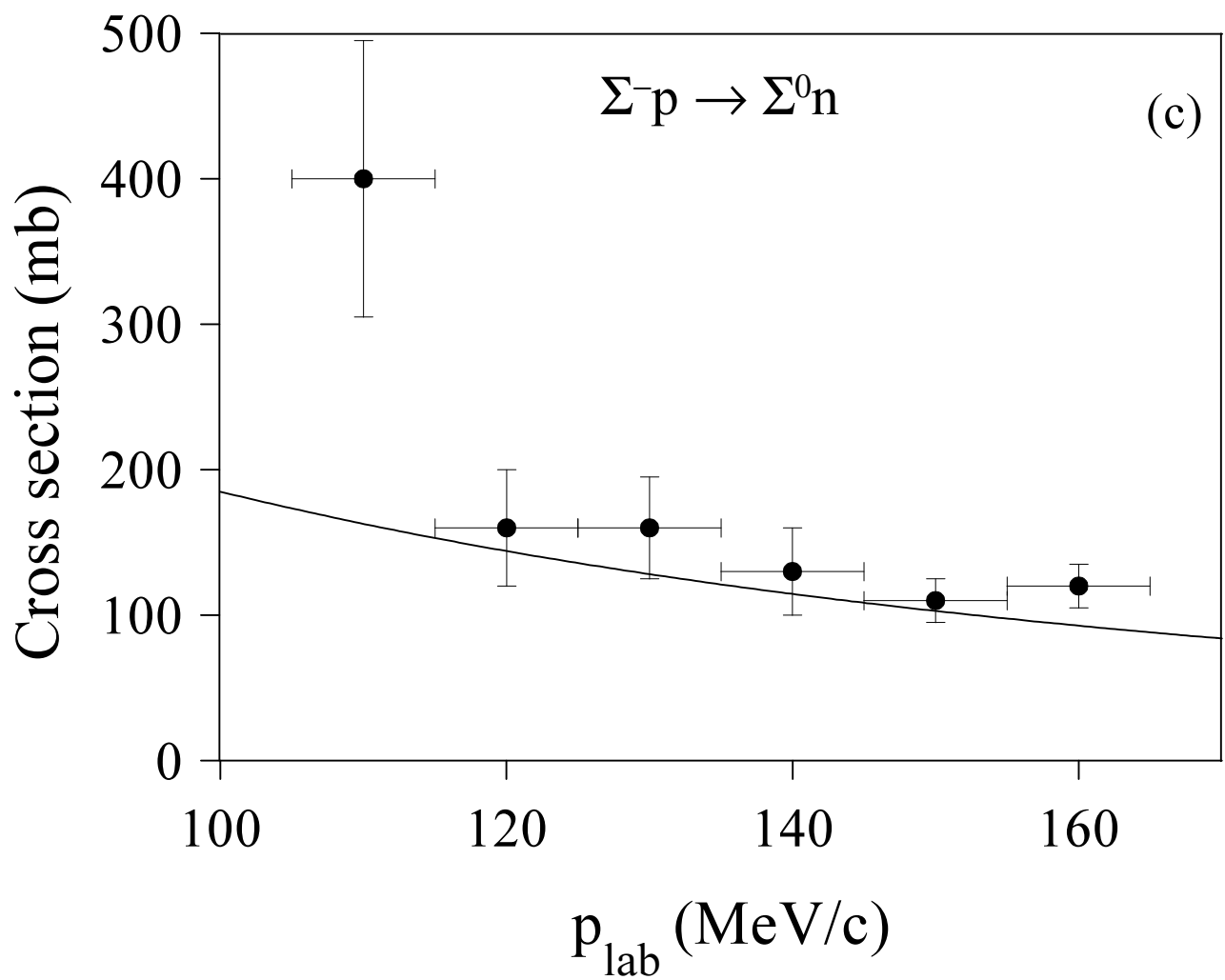
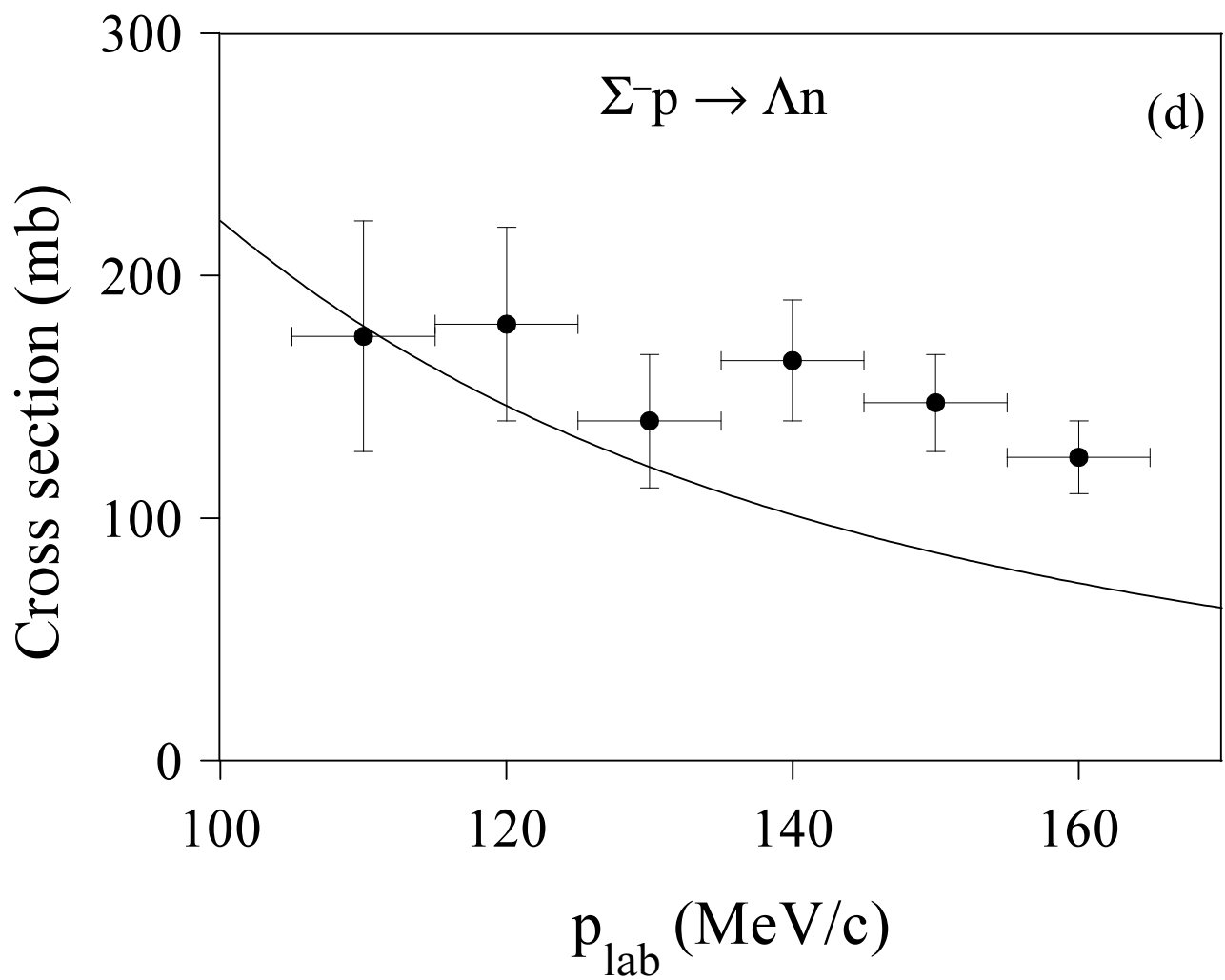
\begin{figure}[tbp]
\caption{Calculated $\Sigma N$ and $\Sigma N \rightarrow \Lambda N$ 
total cross sections compared with
experimental data. Experimental data in (a) and (b) are from Ref. \protect\cite%
{Eis71} and in (c) and (d) from Ref. \protect\cite{Eng66}.}
\label{fig1}
\end{figure}

Our results are plotted in Fig. \ref{fig1}, where a good agreement with the
experimental data is observed. The low-energy parameters for the different
channels are given in Table \ref{t4}. These parameters are complex in the
case of the isospin 1/2 channel due to the fact that the $\Lambda N$
channel is open. A similar agreement for the scattering
cross sections has been obtained in Ref. \cite{Zha97} by means of 
a quark-model based interaction within a resonating group method 
calculation. Our results are also similar to those
obtained by means of effective field theory in next-to-leading order \cite%
{Kor01} or those based on the new Nijmegen soft-core OBE hyperon-nucleon
potential \cite{Rij99}.

\subsection{The three-body system}

As a test of the reliability of our model in the case of the three-baryon
system we solved the $NNN$ bound-state problem. We found a triton binding
energy of 6.90 MeV. For comparison, we notice that the triton binding energy
for the Reid-soft-core potential in the truncated $T-$matrix approximation is
6.58 MeV \cite{Mot99}.

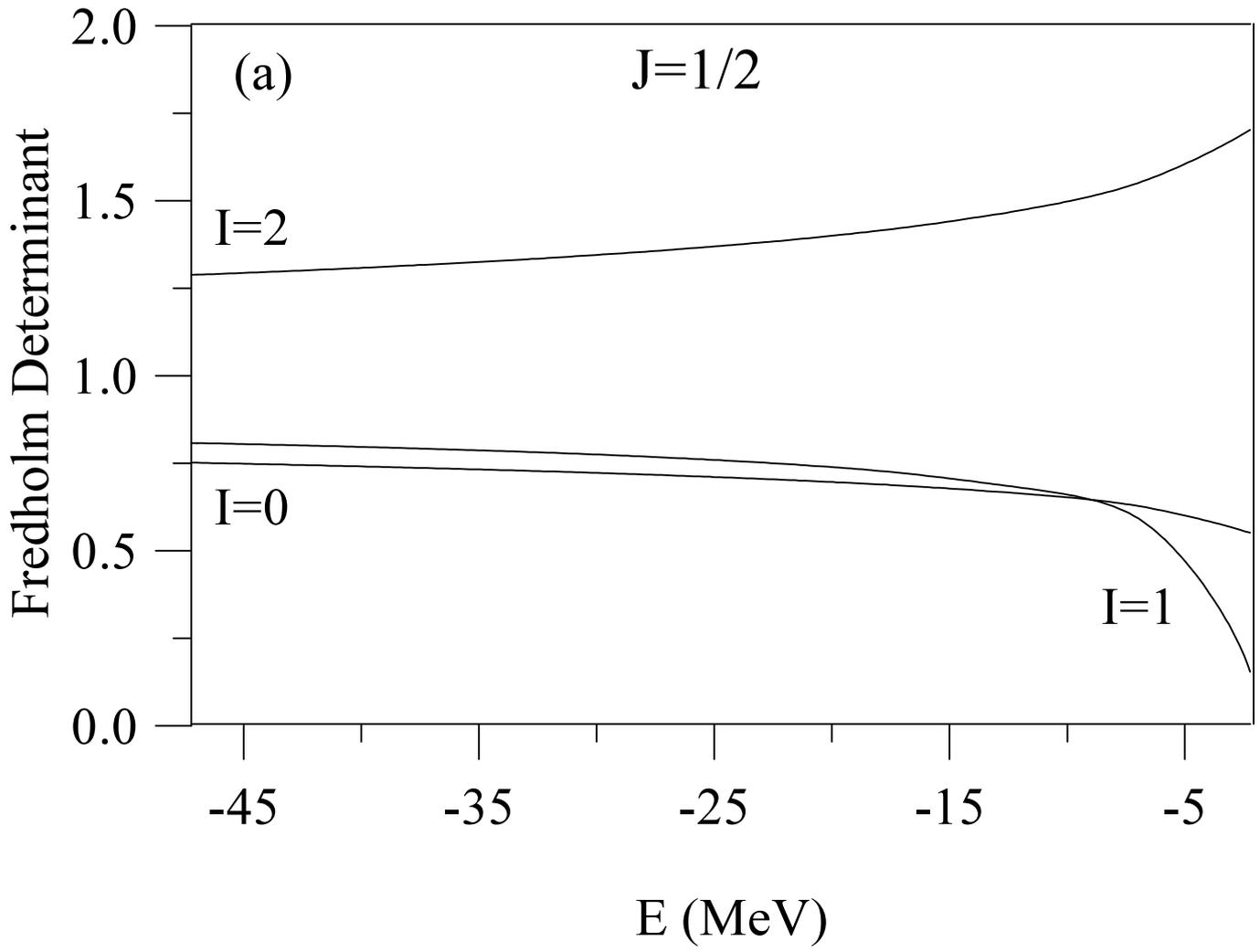
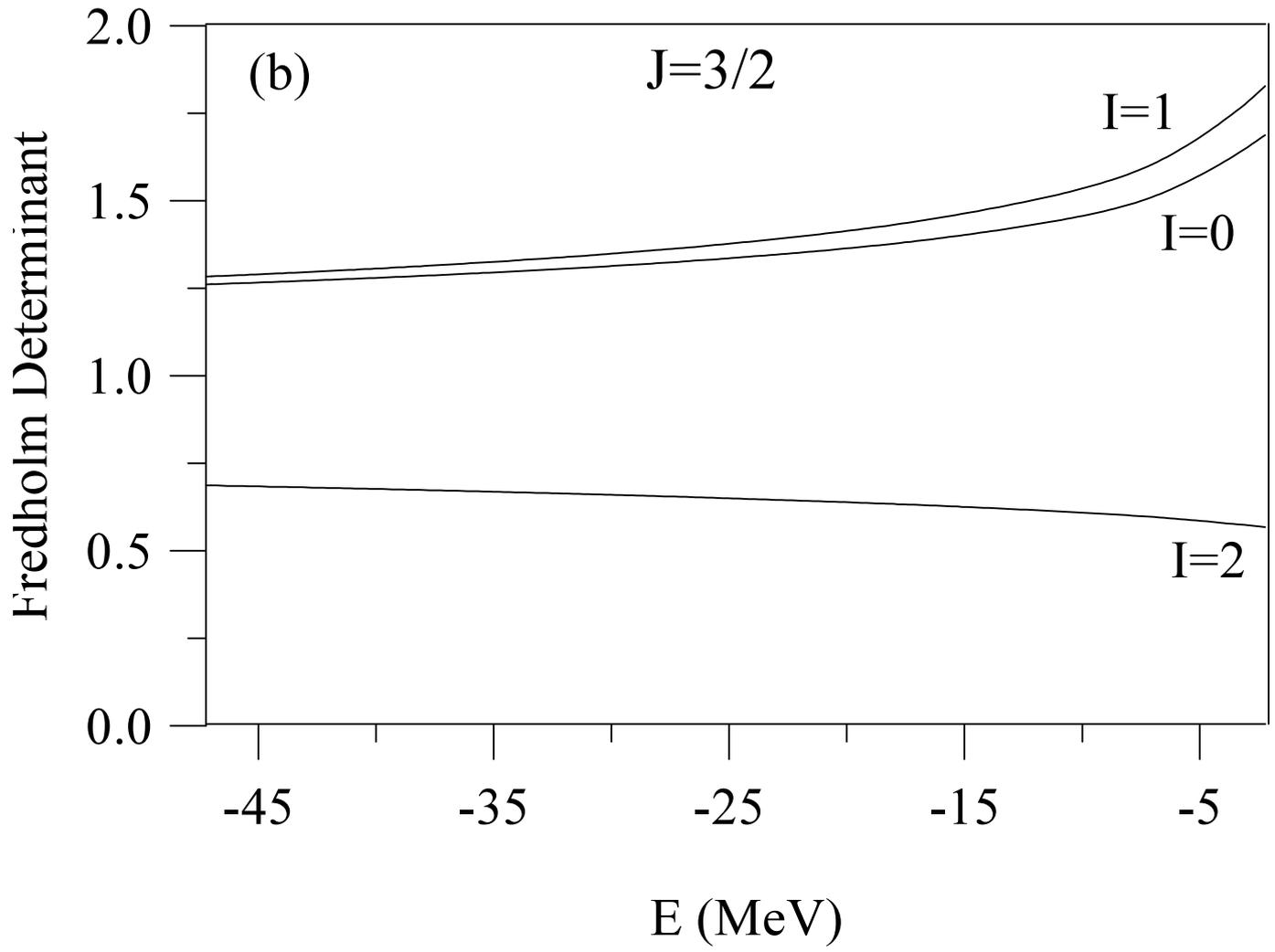
\begin{figure}[tbp]
\caption{Fredholm determinant for (a) $J=1/2$ and (b) $J=3/2$ $\Sigma NN$
channels for the model giving the $\Sigma N$ total cross sections of Fig. 
\protect\ref{fig1}. The $\Sigma d$ continuum starts at $E=-2.225$ MeV, the
deuteron binding energy obtained within our model.}
\label{fig2}
\end{figure}

In Fig. \ref{fig2} we have plotted the Fredholm determinant of the $\Sigma NN
$ system for the three isospin channels with $J=1/2$ and $J=3/2$ calculated
as explained in Sec. \ref{II.C}. As can be
seen there are no bound states. The $J=3/2$ channels are either repulsive or
they do not show any structure, as it is the case of the $I=2$ channel that
remains always flat. For the $J=1/2$ case the $I=2$ channel is repulsive,
while the $I=1$ and $I=0$ are attractive, the $I=1$ being always more
attractive than the $I=0$. If the attraction of the model is increased both
channels develop bound states (the energy ordering between them being
preserved), while all the others remain repulsive, what points out to a
resonance close above the three-body threshold. This is illustrated in Fig. 
\ref{fig3} where we plotted the Fredholm determinant in a model with more 
attraction ($b_s=0.6$ fm), which therefore would not reproduce the $\Sigma N$
scattering cross sections of Fig. \ref{fig1}. As can
be seen the $I=2$, $J=1/2$ case remains equally repulsive while the $I=1$
presents a bound state near threshold. For the $I=0$ state a
resonance behavior close above the three-body threshold is deduced. This
shows that the ordering of the $I=0$ and $I=1$ states with $J=1/2$ is preserved,
the $I=1$ channel being always the lowest state. The order of the two
attractive channels can be easily understood looking at Tables \ref{t3} and %
\ref{t4}. All the attractive two-body channels in the $NN$ and $\Sigma N$
subsystems contribute to the $(I,J)=(1,1/2)$ $\Sigma NN$ state (the $\Sigma N
$ channels $^{3}S_{1}(I=1/2)$ and $^{1}S_{0}(I=3/2)$ and the $^{3}S_{1}(I=0)$
$NN$ channel), while the $(I,J)=(0,1/2)$ state do not present contribution
from two of them, the $^{1}S_{0}(I=3/2)$ $\Sigma N$ and specially the
$^{3}S_{1}(I=0)$ $NN$  deuteron channel. Actually, the $NN$ deuteron-like
contribution plays an essential role in the binding of the triton \cite{Val05}
and hypertriton \cite{Miy95}. In this last case the presence of the $\Lambda$
has the effect of reducing the $NN$ attraction with respect to the deuteron
case but the $\Lambda \leftrightarrow \Sigma$ conversion compensates this reduction and
binds the system. 

In Fig. \ref{fig4} we have plotted the real part of the Fredholm determinant 
of the $\Sigma NN - \Lambda NN$ system for the three isospin channels 
with $J=1/2$ and $J=3/2$ calculated as explained in Sec. \ref{II.D}.  
The imaginary parts are very small and
uninteresting except for the calculation of the widths as we will
see later. As can be seen the inclusion of the $\Lambda NN$
channels does not modify the order of the states, giving values
for the Fredholm determinant very close to the ones of the previous
model (Fig. \ref{fig2}). This shows that the effect of the coupling to
the $\Lambda$ channels is very small at the three-body level once
the coupling to the $\Lambda$ is included at the two-body level.

The pattern of our results coincides exactly with the observations in the 
$^{4}{\rm He}(K_{{\rm stopped}}^{-},N)$ reactions. In particular, we find
only two attractive $S-$wave channels, with the isospin and energy 
ordering corresponding to the experimental $S^{0}(3115)$ and
$S^{+}(3140)$ states. We predict for them $J^{\pi }=1/2^{+}$.

Let us remind that the understanding of these states as deeply bound kaonic nuclear systems 
\cite{Aka02} would assign the quantum numbers $J^{\pi }=3/2^{+}$, $I=0$
for the $S^{0}(3115)$ and $J^{\pi }=1/2^{-}$, $I=1$ for the $S^{+}(3140)$.
If some relativistic effects and a medium-enhanced
$\overline K N$ and spin-orbit interactions are taken 
into account, the ordering of the isospin channels
is reversed to $J^{\pi }=3/2^{+}, I=1$ 
and $J^{\pi }=1/2^{+}, I=0$. The $SU(3)$ multiskirmion 
description \cite{Sca00} finds $J^{\pi }=1/2^{+}$ for both states, 
but the opposite ordering between the isospin
states with respect to our results and experiment. 
The nona-quark study of Ref. \cite{Mae05} makes use of a
Gell-Mann-Okubo like mass formula to study the spectrum of $S=-1,-2,-3$
nona-quark states. The color magnetic interaction between quarks, together
with the antisymmetrization of the wave function, favors small multiplets in
flavor and spin which gives a natural explanation for the $I=1$ state being
the lowest state among the $S=-1$ tribaryons with $J=1/2$. This leads to the
natural explanation that the $I=1$ state could be a member of the $27-$plet
with $J^{\pi }=1/2$, and the $I=0$ state may be a member of the $10^{\ast}-$plet
with $J^{\pi }=3/2$. However other possible classifications that may
give rise to $J^{\pi }=3/2$ for the $S^{0}(3115)$ and the $S^{+}(3140)$ were
also discussed. 

\subsection{Calculation of the widths}

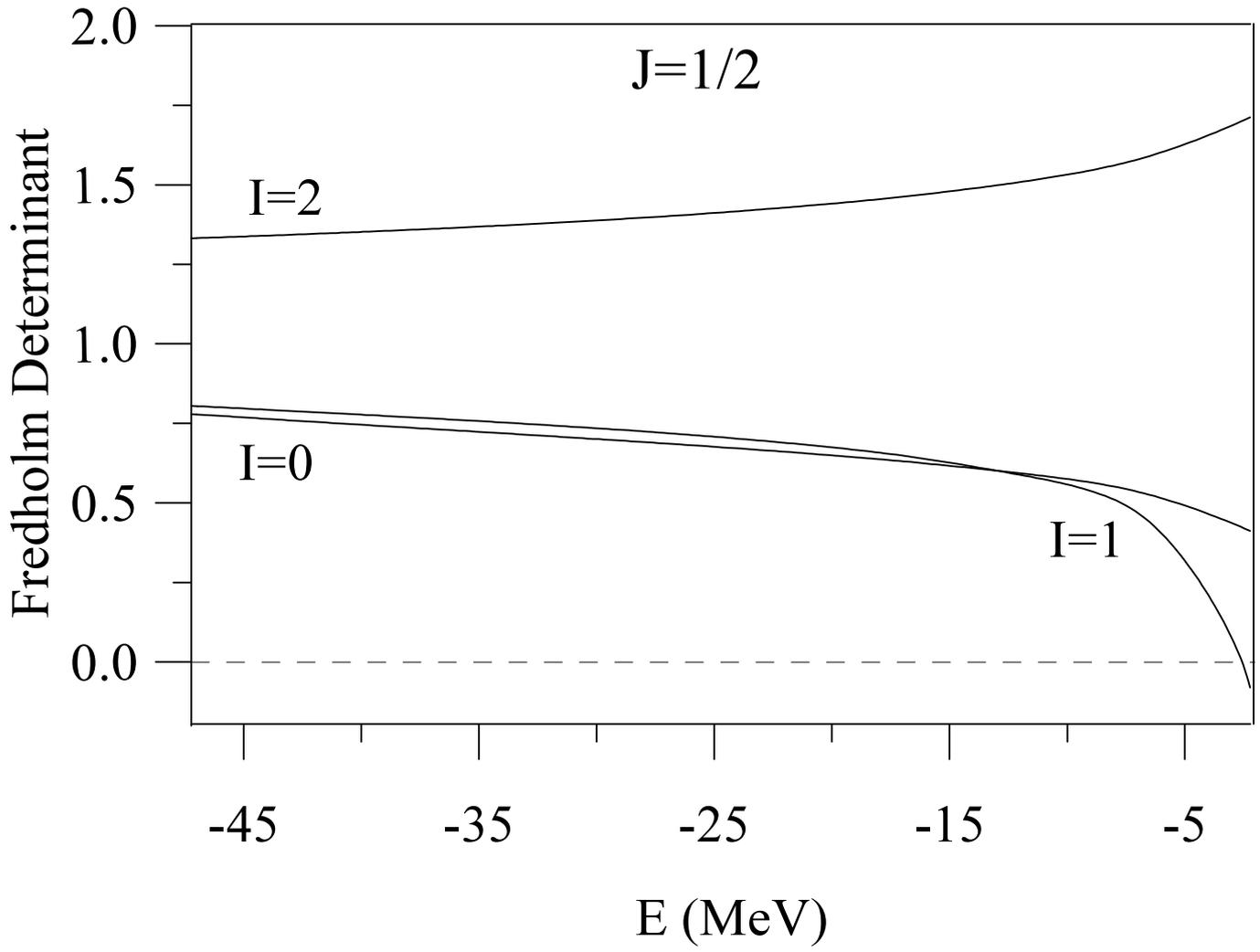
\begin{figure}[tbp]
\caption{$J=1/2$ $\Sigma NN$ Fredholm determinant for a model with increased 
attraction as explained in the text.}
\label{fig3}
\end{figure}

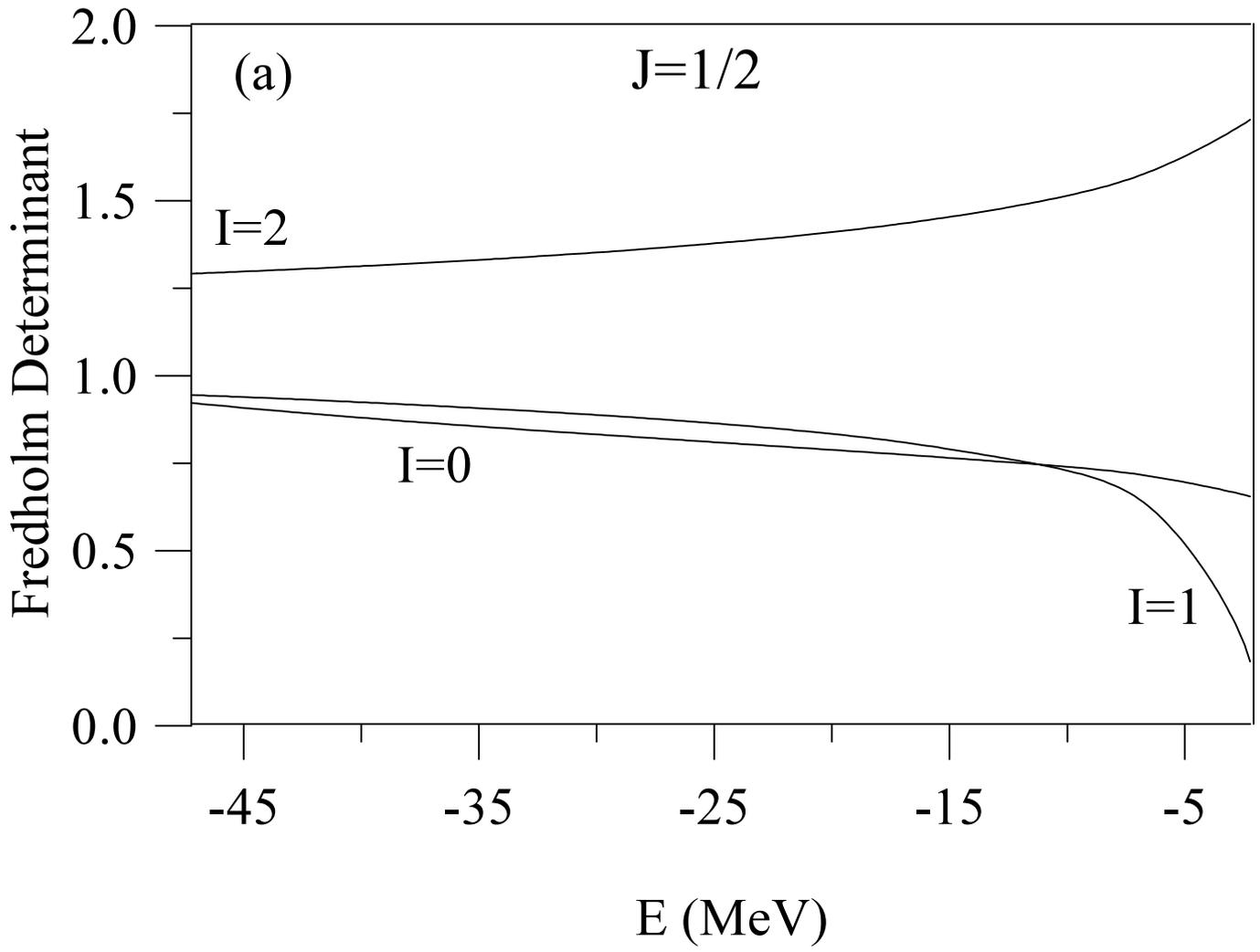
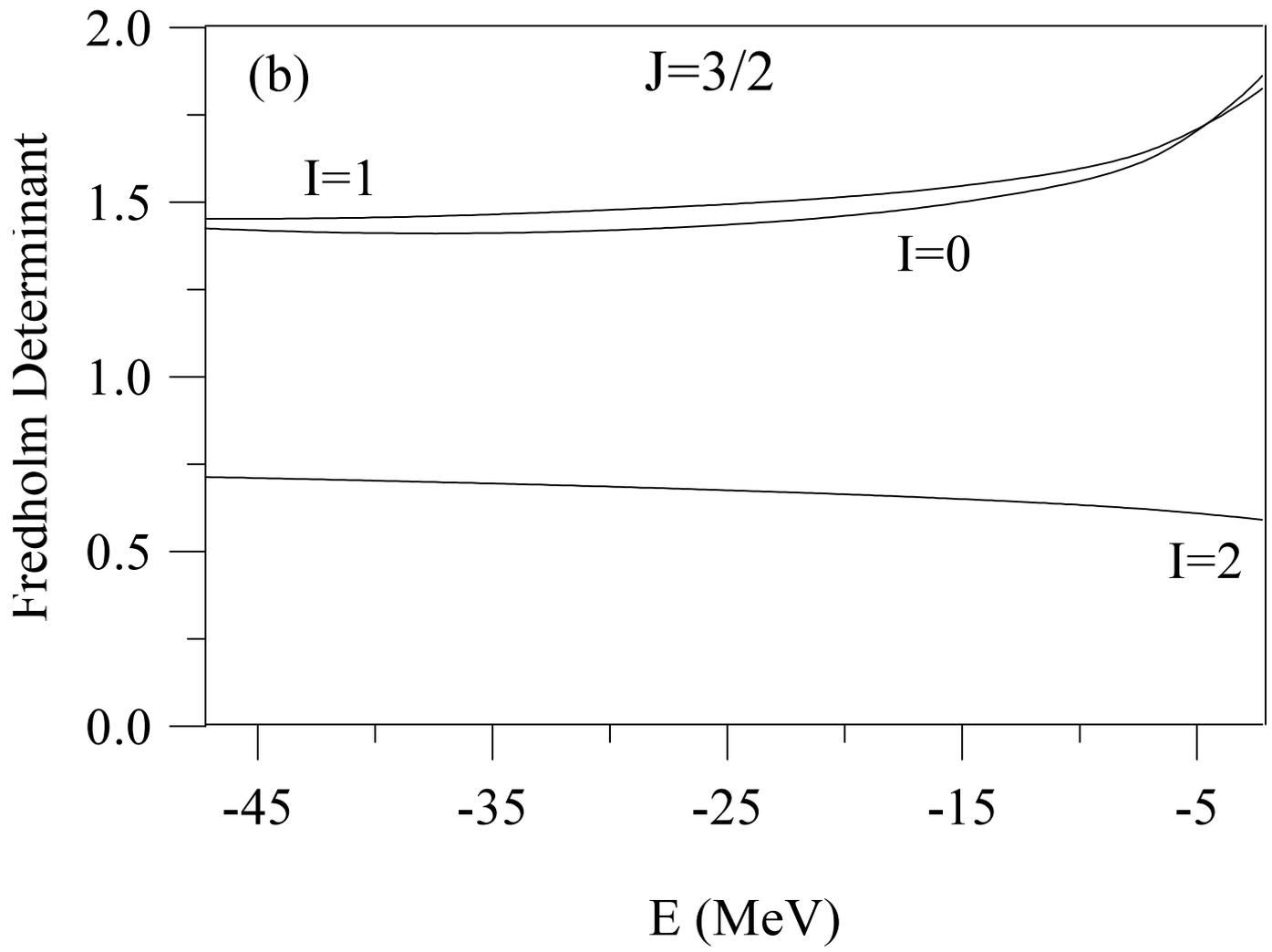
\begin{figure}[tbp]
\caption{Real part of the Fredholm determinant for (a) $J=1/2$ and (b) $J=3/2$ 
$\Sigma NN - \Lambda NN$
channels for the model giving the $\Sigma N$ total cross sections of Fig. 
\protect\ref{fig1}. The $\Sigma d$ continuum starts at $E=-2.225$ MeV, the
deuteron binding energy obtained within our model.}
\label{fig4}
\end{figure}

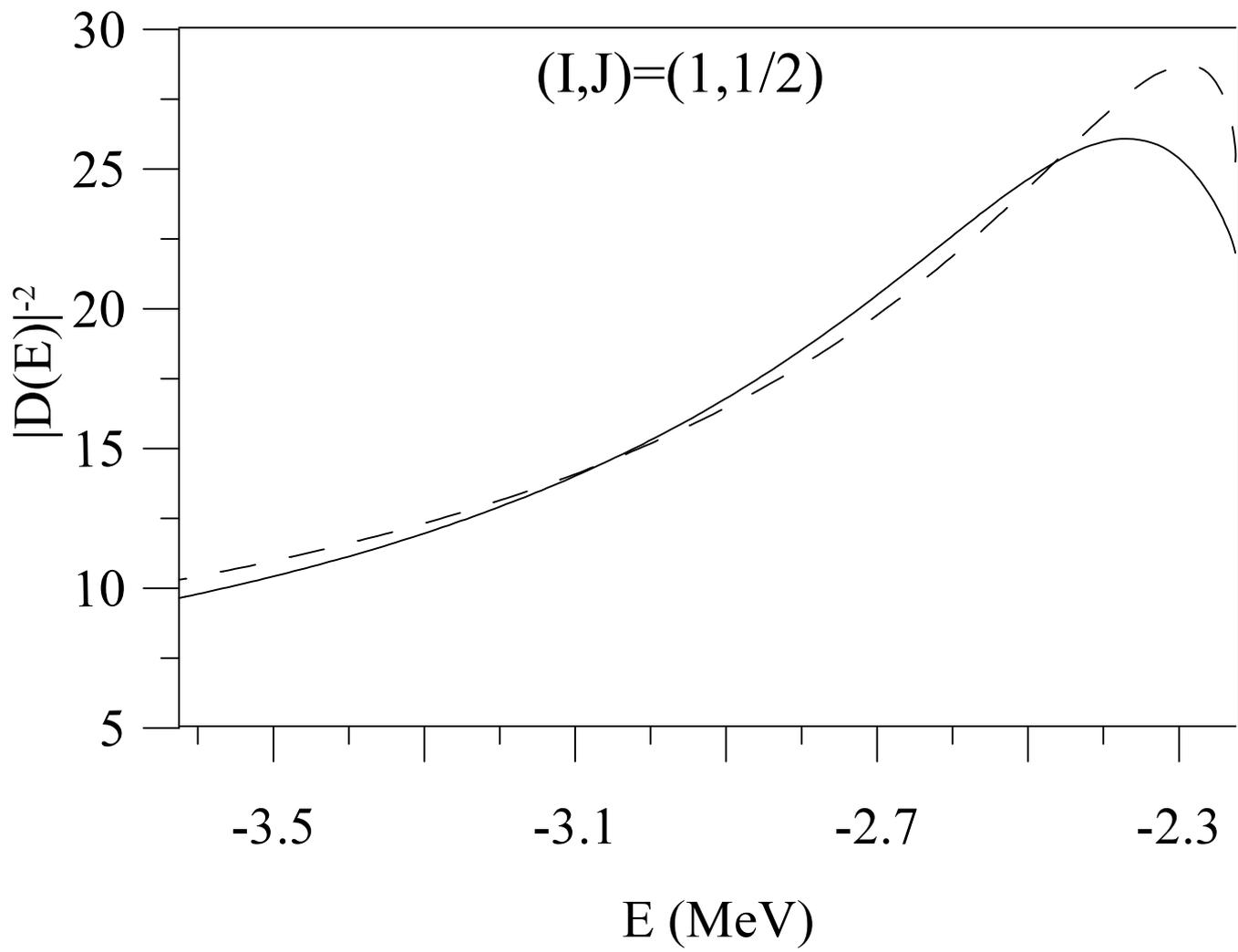
\begin{figure}[tbp]
\caption{Inverse of the square of the Fredholm determinant, 
$1/\mid D(E)\mid^2$, for the bound state case of 
Fig. \protect\ref{fig3}, $(I,J)=(1,1/2)$, using the full 
propagator $G_{\Lambda}(E;p_\Lambda)$ in Eq. 
(\protect\ref{far1}) (dashed line) and considering also
the $\Lambda NN$ channels (solid line).}
\label{fig5}
\end{figure}

In Fig. \ref{fig3} we have shown the Fredholm determinant in the case when 
there is a bound state in the $(I,J)=(1,1/2)$ channel. As discussed in
subsection \ref{II.C}, the Fredholm determinant is real since we have dropped 
the imaginary part of the propagator $G_{\Lambda}(E;p_\Lambda)$
in Eq. (\ref{far1}). Near the bound state the Fredholm determinant has the
form $D(E)=C(E-E_0)$ where $C$ is a constant and $E_0$ is the energy 
of the bound state. If we now repeat the calculation using the full 
propagator $G_{\Lambda}(E;p_\Lambda)$ given by Eq. (\ref{far1}) the Fredholm
determinant becomes complex. Near the bound state it has the form
$D(E)=C[(E-E_0)+i\Gamma]$ so that $1/\mid D(E)\mid^2$ has the resonant 
shape

\begin{equation}
{1\over \mid D(E)\mid^2}={1\over \mid C\mid^2[(E-E_0)^2+\Gamma^2]},
\end{equation}
which is also the shape exhibited by the cross section near a 
resonance ($\sigma(E)\propto 1/\mid D(E)\mid^2$).
In Fig. \ref{fig5} we show $1/\mid D(E)\mid^2$,
from which we extract $\Gamma=0.3$ MeV for the model without $\Lambda NN$
channels and $\Gamma=0.5$ MeV for the model with $\Lambda NN$ channels.
This state lies 80 MeV above the $\Lambda NN$ threshold while the 
observed tribaryons lie at 120 MeV and 140 MeV, respectively, above the
$\Lambda NN$ threshold.

Since the state of Fig. \ref{fig3} has a width of less than 1 MeV one can 
reasonably expect that the observed states which lie 40 and 60 MeV
above it will have somewhat larger widths but certainly in agreement
with the experimental result $\Gamma < 21$ MeV.

\section{Conclusions}

We have studied the bound-state solutions of the $\Sigma NN$ system by means
of interactions derived from a constituent quark cluster model. The two-body
interactions correctly reproduce the low-energy observables of the 
$NN$ and $\Sigma N$ subsystems.
We have not found any $\Sigma NN$ bound state. However, 
our results show that there are only two attractive $S-$wave
channels, they are the $(I,J)=(1,1/2)$ and $(0,1/2)$, with a
resonance close above the three-body threshold. The channel with $I=1$
is always more attractive than that with $I=0$. The isospin quantum
numbers and the energy ordering correspond exactly to the recently measured
strange tribaryons from the $^{4}{\rm He}(K_{{\rm stopped}}^{-},N)$
reactions in the KEK PS E471 experiment. 
We predict quantum numbers $J^{\pi }=1/2^{+}$ and small widths
for the two reported strange tribaryon resonances.  The awaited
experimental determination of $J^\pi$ can serve as a stringent
test of our model dynamics against others.

\acknowledgements

This work has been partially funded by Ministerio de Educaci\'{o}n y Ciencia
under Contract No. FPA2004-05616, by Junta de Castilla y Le\'{o}n
under Contract No. SA-104/04, by COFAA-IPN (M\'{e}xico) and by Oficina
de Ciencia y Tecnolog\'{\i}a de la Comunidad Valenciana, Grupos 03/094
and GV05/276.

\begin{table}[tbp]
\caption{$\Sigma N$ channels $(\ell_\Sigma,s_\Sigma)$ and 
$\Lambda N$ channels $(\ell_\Lambda,s_\Lambda)$ that contribute
to a given $\Sigma N$ state 
with isospin $i$ and total angular momentum $j$.}
\label{t1}%
\begin{tabular}{cccccc}
& $i$ & $j$ & $(\ell_\Sigma,s_\Sigma)$ & $(\ell_\Lambda,s_\Lambda)$ &  \\ 
\tableline 
& 3/2 & 0 & (0,0) & & \\ 
& 3/2 & 1 & (0,1),(2,1) & & \\ 
& 1/2 & 0 & (0,0) & (0,0) & \\ 
& 1/2 & 1 & (0,1),(2,1) & (0,1),(2,1) & \\ 
\end{tabular}
\end{table}

\begin{table}[tbp]
\caption{$NN$ channels $(\ell_N,s_N)$ and $\Delta N$ channels 
$(\ell_\Delta,s_\Delta)$ that are coupled together in the 
$^3S_1-\, ^{3} \! D_1$, and $^1S_0$ $NN$ states.}
\label{t2}%
\begin{tabular}{ccccc}
$NN$ state & $i$ & $j$ & $(\ell_N,s_N)$ & $(\ell_\Delta,s_\Delta)$ \\ 
\tableline $^3S_1-\, ^{3} \! D_1$ & 0 & 1 & (0,1),(2,1) &  \\ 
$^1S_0$ & 1 & 0 & (0,0) & (2,2) \\ 
\end{tabular}
\end{table}

\begin{table}[tbp]
\caption{Two-body $\Sigma N$ channels $(i_\Sigma,s_\Sigma)$,
$\Lambda N$ channels $(i_\Lambda,s_\Lambda)$,
$NN$ channels with $\Sigma$ spectator $(i_{N(\Sigma)},s_{N(\Sigma)})$, and
$NN$ channels with $\Lambda$ spectator $(i_{N(\Lambda)},s_{N(\Lambda)})$ that contribute to
a given $\Sigma NN - \Lambda NN$ state with total isospin $I$ and spin $J$.}
\label{t3}
\begin{tabular}{cccccc}
$I$ & $J$ & $(i_\Sigma,s_\Sigma)$  & $(i_\Lambda,s_\Lambda)$ 
& $(i_{N(\Sigma)},s_{N(\Sigma)})$ & $(i_{N(\Lambda)},s_{N(\Lambda)})$ \\ 
\tableline 
0 & 1/2 & (1/2,0),(1/2,1)  & (1/2,0),(1/2,1)  & (1,0) & (0,1) \\ 
1 & 1/2 & (1/2,0),(3/2,0),(1/2,1),(3/2,1)  & (1/2,0),(1/2,1) & (0,1),(1,0) &
(1,0)  \\ 
2 & 1/2 & (3/2,0),(3/2,1) & & (1,0) &  \\ 
0 & 3/2 & (1/2,1)  & (1/2,1) & & (0,1) \\ 
1 & 3/2 & (1/2,1),(3/2,1)  & (1/2,1) & (0,1) & \\ 
2 & 3/2 & (3/2,1) &  & & \\ 
\end{tabular}
\end{table}

\begin{table}[tbp]
\caption{Low-energy scattering parameters (in fm) of the $\Sigma N$ $^1S_0$
and $^3S_1$ channels for the states with total isospin $i=1/2$ and $i=3/2$.}
\label{t4}
\begin{center}
\begin{tabular}{ccccc}
& \multicolumn{2}{c}{$^1S_0$} & \multicolumn{2}{c}{$^3S_1$} \\ 
\cline{2-3}\cline{4-5}
& $a_s$ & $r_s$ & $a_t$ & $r_t$ \\ \hline
$i=1/2$ & $-1.24 +i 0.08$ & $-0.80 -i 0.33$ & $4.65 +i 4.22$ & $3.13 -i 0.43$\\ 
$i=3/2$ & $3.16$ & $4.78$ & $-0.72$ & $-0.63$ \\ 
\end{tabular}
\end{center}
\end{table}

\end{document}